\documentclass[aps,prl,showpacs,twocolumn,superscriptaddress]{revtex4}
\usepackage{bm}
\usepackage{epsfig}
\usepackage{times}
\usepackage{amssymb}
\usepackage{subfigure}

\begin{document}
\title{Exact Nondipole Kramers-Henneberger Form of the Light-Atom Hamiltonian:
An Application to Atomic Stabilization and Photoelectron Energy
Spectra}

\author{M. F\o rre}
\affiliation{Department of Physics and Technology, University
  of Bergen, N-5007 Bergen, Norway}
\author{S. Selst\o}
\affiliation{Department of Physics and Technology, University
  of Bergen, N-5007 Bergen, Norway}
\author{J. P. Hansen}
\affiliation{Department of Physics and Technology, University
  of Bergen, N-5007 Bergen, Norway}
\author{L. B. Madsen}
\affiliation{Department of Physics and Astronomy, University
  of Aarhus, 8000 Aarhus, Denmark}

\begin{abstract}
The exact nondipole minimal-coupling Hamiltonian for an atom
interacting with an explicitly time- and space-dependent laser
field is transformed into the rest frame of a classical free
electron in the laser field, i.e., into the Kramers-Henneberger
frame. The new form of the Hamiltonian is used to study  nondipole
effects in the high-intensity, high-frequency regime. Fully three
dimensional nondipole {\it ab initio} wavepacket calculations show
that the ionization probability may decrease for increasing field
strength. We identify a unique signature for the onset of this
dynamical stabilization effect in the photoelectron spectrum.
\end{abstract}

%42.50.Hz Strong-field excitation of optical transitions in quantum
%systems; multi-photon processes; dynamic Stark shift (for
%multiphoton ionization and excitation of atoms and molecules, see
%32.80.Rn, and 33.80.Rm, respectively)
%
%32.80.Rm Multiphoton ionization and excitation to highly excited
%states (e.g., Rydberg states)
%
%32.80.Fb Photoionization of atoms and ions

\maketitle

The general field of laser-matter interactions is characterized by
impressive progress in light-source technology. Light sources with
pulses of shorter and shorter duration and ever increasing
intensities are being developed. Pulses containing only a few
cycles and with a duration of less than 10 fs are now commercially
available~\cite{Brabec00}. Intensities of 10$^{14}$ W/cm$^{2}$ are
routinely provided, and intensities two orders of magnitude
higher, reaching the field strength of the Coulomb interaction in
atoms and molecules, are not unusual. Femtosecond laser pulses
have been used to produce coherent extreme-ultraviolet pulses of
attosecond duration, and the expression ``attosecond
metrology''~\cite{Hentschel01} was coined for the investigation of
matter with such short pulses~\cite{Kienberger04}. Other
developments include the large-scale intense free-electron
laser projects at DESY (Hamburg, Germany) and SLAC (Stanford,
USA). The TESLA test facility in Hamburg has begun operation in
the far-ultraviolet regime and, e.g., a study of the interaction
of intense soft X-rays with atom clusters was
reported~\cite{Wabnitz02}. The clusters absorbed energy much more
efficiently than anticipated from existing models, and the
physical mechanism responsible for the excess in the absorbed
energy is currently subject to some
controversy~\cite{Santra03Siedschlag04}.

Typically the laser-atom interaction is described in the dipole
approximation where several equivalent formulations exist; the
most popular ones being the velocity gauge, the length gauge and
the Kramers-Henneberger frame~\cite{CohentannoudjiLoudon}. It is,
however, clear that the new light sources alluded to above pave
the way for studies of atomic and molecular systems under extreme
nonperturbative conditions~\cite{Meharg}. In the case of atoms
interacting with light from the vacuum-ultra-violet free-electron
laser the dipole approximation cannot be expected to be
valid~\cite{Lugovskoy}. Thus, motivated by the need to include the
full ${\bm k} \cdot {\bm r}$-term in the description of the
light-matter interaction, we here revisit the question of
equivalent formulations of electrodynamics.

We transform the exact nondipole minimal-coupling Hamiltonian for
an atom in an explicitly time- and space-dependent field into the
rest frame of a classical free electron in the laser field. In the
dipole approximation, this frame is known as the
Kramers-Henneberger frame~\cite{PKH}. Our transformed exact
nondipole Hamiltonian takes a simple form and is very useful for
the discussion of strong-field dynamics. We apply it to the study
of H in the high-intensity, high-frequency regime, and confirm the
phenomenon of atomic stabilization, i.e., the possibility of
having a decreasing ionization probability/rate with increasing
intensity (for reviews see, e.g., \cite{GavrilaPopov}). Most
importantly, we point out that the onset of the dynamic
stabilization can be directly observed from electron energy
spectra. [Atomic units (a.u.) with $m_e=e=\hbar = 1$ are used
throughout. All derivations are straightforwardly generalized to
atoms and molecules involving more electrons.]

The minimal coupling scheme determines the Hamiltonian for a
charged particle in an electromagnetic field through the vector
potential ${\bm A}(\eta)$ with $\eta \equiv \omega t - {\bm k} \cdot
{\bm r}$, and ${\bm k}$  the wave number. The scheme implies that
the canonical momentum is obtained by  ${{\bm p}} \rightarrow {\bm
p} - q {\bm A}$ and for an electron of charge $q=-1$ in atomic
units, we have ${\bm p} +{\bm A}$, and the time-dependent
Schr\"{o}dinger equation reads
\begin{equation} \label{v-form}
i{\partial}_t \Psi_v({\bm r},t)=\left[ ({\bm p}+{\bm A}(\eta))^2+
V({\bm r})\right]\Psi_v({\bm r},t),
\end{equation}
where the subscript $v$ refers to the velocity gauge. The
advantage of this formulation is that the spatial dependence of
the field is explicitly accounted for through its presence in the
vector potential. A disadvantage is that the interaction is not
expressed in terms of the physical $E$- and $B$-fields. Also
numerically, the evaluation of the action of the ${\bm A} \cdot
{\bm p}$-term can be quite involved unless a diagonal
representation of $\Psi_v$ with respect to this operator is
applied. Until now only the alternative multipole formulation of
Power-Zienau-Woolley~\cite{CohentannoudjiLoudon,PZW} has, in
principle,  kept the spatial dependence to all orders. The
multipolar form represents the interaction in terms of the
physical fields and the electron coordinate ${\bm r}$, but, as the
name suggests, it is inherently designed to provide an expansion
of the light-matter interaction, and consequently very impractical
if one wishes to retain ${\bm k} \cdot {\bm r}$ to all orders.

Here, we transform the Schr\"{o}dinger equation into a new form by
applying a nondipole Kramers-Henneberger transformation. Let
\begin{equation}\label{trans}
\Psi_{K\! H}=U \Psi_v = \exp[i \, \mbox{\boldmath $\alpha$}(\eta)
\cdot {\bm p}] \Psi_v,
\end{equation}
where
\begin{equation}\label{alpha}
\mbox{\boldmath $\alpha$}(\eta)\equiv\frac{1}{\omega}
\int_{\eta_i}^{\eta} d\eta' \, {\bm A} (\eta')
\end{equation}
represents the quiver motion relative to the laboratory frame of a
classical free electron in the field. The Hamiltonian
corresponding to the new point of view is obtained by taking the
time-derivative on both sides of (\ref{trans}), and by using
(\ref{v-form}) for $\Psi_v$,  we obtain $i\partial_t \Psi_{K\! H}(\bm
r, t) = H_{K\! H}  \Psi_{K\! H}(\bm r, t)$ with
\begin{equation}
\label{HKH-gen} H_{K\! H} = U H_v U^\dagger + i (\partial_t U)
U^\dagger.
\end{equation}
To evaluate the effect of the unitary translation operators in
(\ref{HKH-gen}), we use the operator identity known as the
Baker-Hausdorff lemma~\cite{Sakurai} and take advantage of the
Coulomb gauge restriction $[{\bm p}, {\bm A}] = 0$ and  ${\bm k}
\cdot {\bm A} = 0$.
\begin{figure}
\begin{center}
\epsfysize=5.5cm \epsfbox{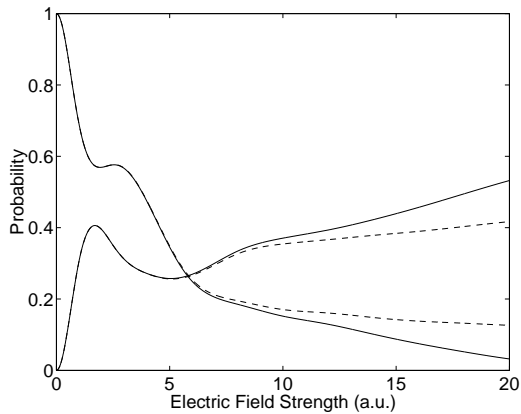}
\end{center}
\caption{Ionization and ground state probability for a
two-dimensional model atom~\cite{Kylstra} in the nondipole (solid
curve) and dipole (dashed curve) descriptions vs.\ electric field
strength for a 5 cycle pulse with $\omega=1$ a.u.. } \label{fig1}
\end{figure}
The resulting Hamiltonian reads
\begin{widetext}
\begin{eqnarray}
\label{HKH} H_{K\!H}=\frac{p^2 + A^2}{2}  +V({\bm r}+
\mbox{\boldmath $\alpha$})  + \frac{k^2}{2}\left( \frac{d
\mbox{\boldmath $\alpha$}}{d\eta} \cdot {\bm p}\right)^2
+\frac{ik^2}{2} \;\frac{d^2 \mbox{\boldmath $\alpha$}}{d \eta^2}\;
\cdot {\bm p} +  \left(\frac{d \mbox{\boldmath $\alpha$}}{d\eta}
\cdot {\bm p}\right) ({\bm k}\cdot {\bm p}),
\end{eqnarray}
\end{widetext}
which holds for a general elliptically polarized
field. Within the dipole approximation ${\bm A}$ and
$\mbox{\boldmath $ \alpha$}$ are space-independent, the last three
terms are absent, and (\ref{HKH}) reduces to the well-known
result~\cite{PKH}. In the nondipole case, the importance of these
terms is readily understood, e.g., in terms of their effect on a
continuum wave function. The two terms proportional to $k^2$ are
of the order of $E_0^2v^2/(\omega^2 c^2)$ and $E_0 v/c^2$,
respectively, whereas the last term is of order $E_0 v^2/(\omega
c)$. We thus see that the effect of the dominant term on a wave
function is reduced by a factor $\sim E_0/(\omega c)$ compared to
the $p^2$-term. The factor $E_0/\omega$ is precisely the quiver
velocity of the electron $v_{\text{quiver}}$, so  we expect that
the last three terms may be neglected as long as $
v_{\text{quiver}} / c \ll 1$. Whenever this condition is
fulfilled, the non-relativistic approach is automatically
justified as well. As it turns out, for the field parameters
considered here, the effect of the nondipole terms is effectively
given by the spatial dependence of the vector potential in the
$A^2$-term.

As a first application of the new form of the Hamiltonian we
consider the interaction with high-intensity, high-frequency
fields. In this so-called stabilization
regime~\cite{GavrilaPopov}, atoms may go through a region of
decreasing ionization for increasing field strength. Stabilization
was experimentally observed with Rydberg atoms~\cite{Boer}. With
the development of new light sources, dynamic stabilization of
ground state atoms is, however, expected to be within experimental
reach in the near future~\cite{dondera}.

Nondipole terms were investigated in approximate ways earlier and
found to have a detrimental effect on the
stabilization~\cite{Bugacov,Kylstra}. The relative role of the
different nondipole terms in (\ref{HKH}) is illustrated in
Fig.~\ref{fig1} for a two-dimensional model atom~\cite{Kylstra}.
The ground state was exposed to a laser pulse propagating in the
$x$ direction and of linear polarization ${\bm u_p}$ along the
$z$-axis corresponding to the vector potential ${\bm
A}(\eta)=\frac{E_0}{\omega}f(\eta) \sin(\eta\ + \phi) {\bm u_p}$
with $f(\eta)$ the envelope, $E_0$ the electric field amplitude
and $\phi$ a phase that ensures that the vector potential
represents a physical field~\cite{Lars2}. The wave function was
propagated on a Cartesian grid by means of the split-step operator
technique \cite{HermannFleck}. A 5-cycle laser pulse with central
frequency $\omega=1$ a.u. ($46$ nm) corresponding to the pulse
duration $T=760$ as, and with carrier-envelope
$f(\eta)=\sin^2\left(\frac{\pi\eta}{\omega T}\right)$, was
employed. The intensity range was set to $0<I_0<1.4\times10^{19}$
W/cm$^{2}$. The population not accounted for in Fig.~\ref{fig1},
is left in excited states. The effects of the last three terms in
the Hamiltonian as well as the spatial dependence of the quiver
amplitude $\mbox{\boldmath $ \alpha$}({\bm r},t)$ cannot be
resolved on the scale of Fig.~\ref{fig1}. Hence, the nondipole
effect observed for field strengths greater than 7 a.u. are
exclusively related to the spatial dependence of the $A^2$-term.

We have, accordingly, justified that for the parameters under
concern, it is a very accurate approach to apply the Hamiltonian
(\ref{HKH}), neglecting the last three additional kinetic energy
terms arising from the transformation (\ref{trans}), to a fully
three-dimensional study of ionization of a real ground state atom
by intense short wave light field beyond the dipole approximation.
We consider H(1s) exposed to 5-cycle laser pulses in the
attosecond range with central frequencies $\omega=1$ a.u. and
$\omega=2$ a.u.. The time-dependent Schr{\"o}dinger equation is
solved numerically based on a split-step operator approximation on
a spherical grid as detailed elsewhere~\cite{JanPetter}.
\begin{figure}
\begin{center}
\epsfysize=5.5cm \epsfbox{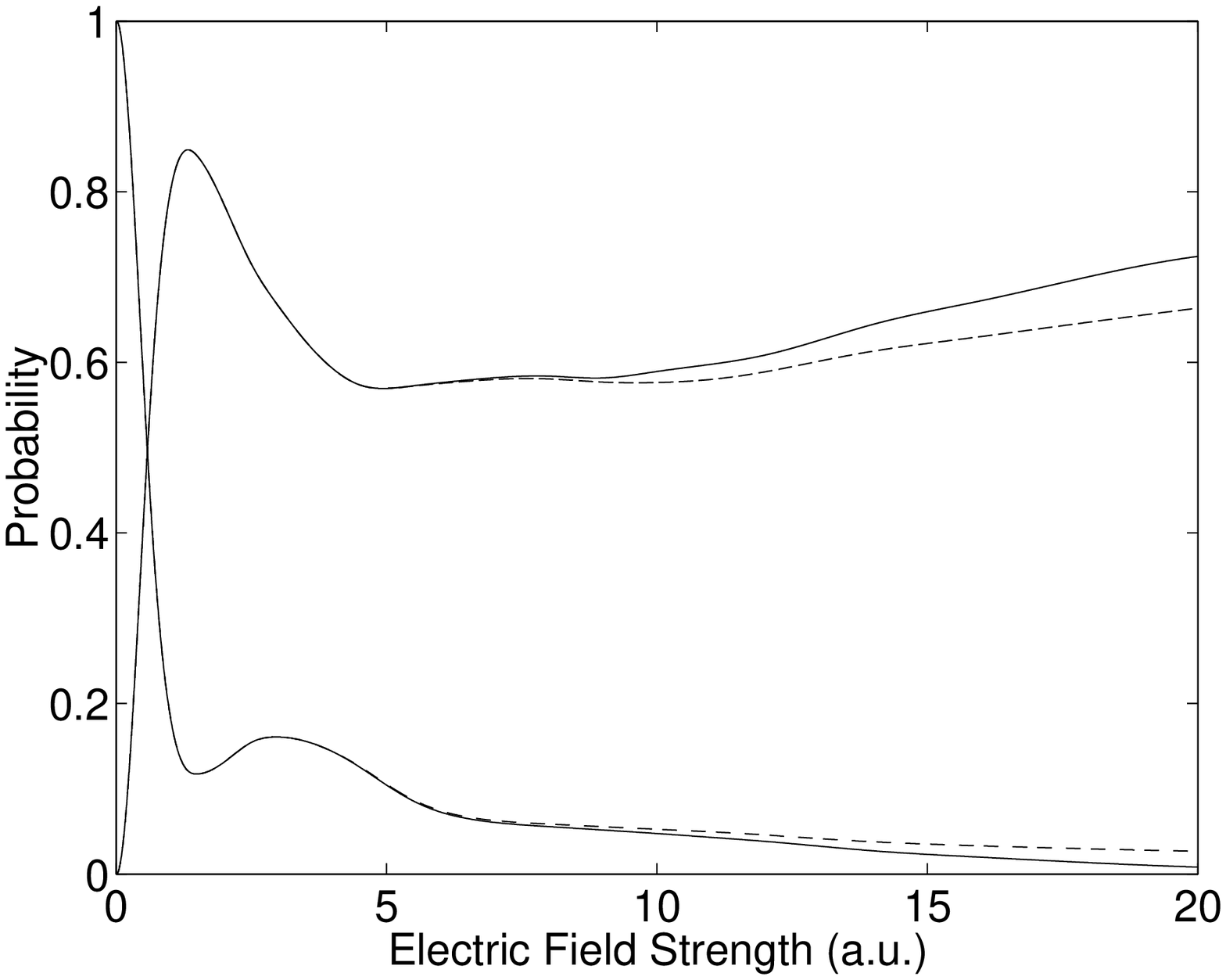}
\epsfysize=5.5cm \epsfbox{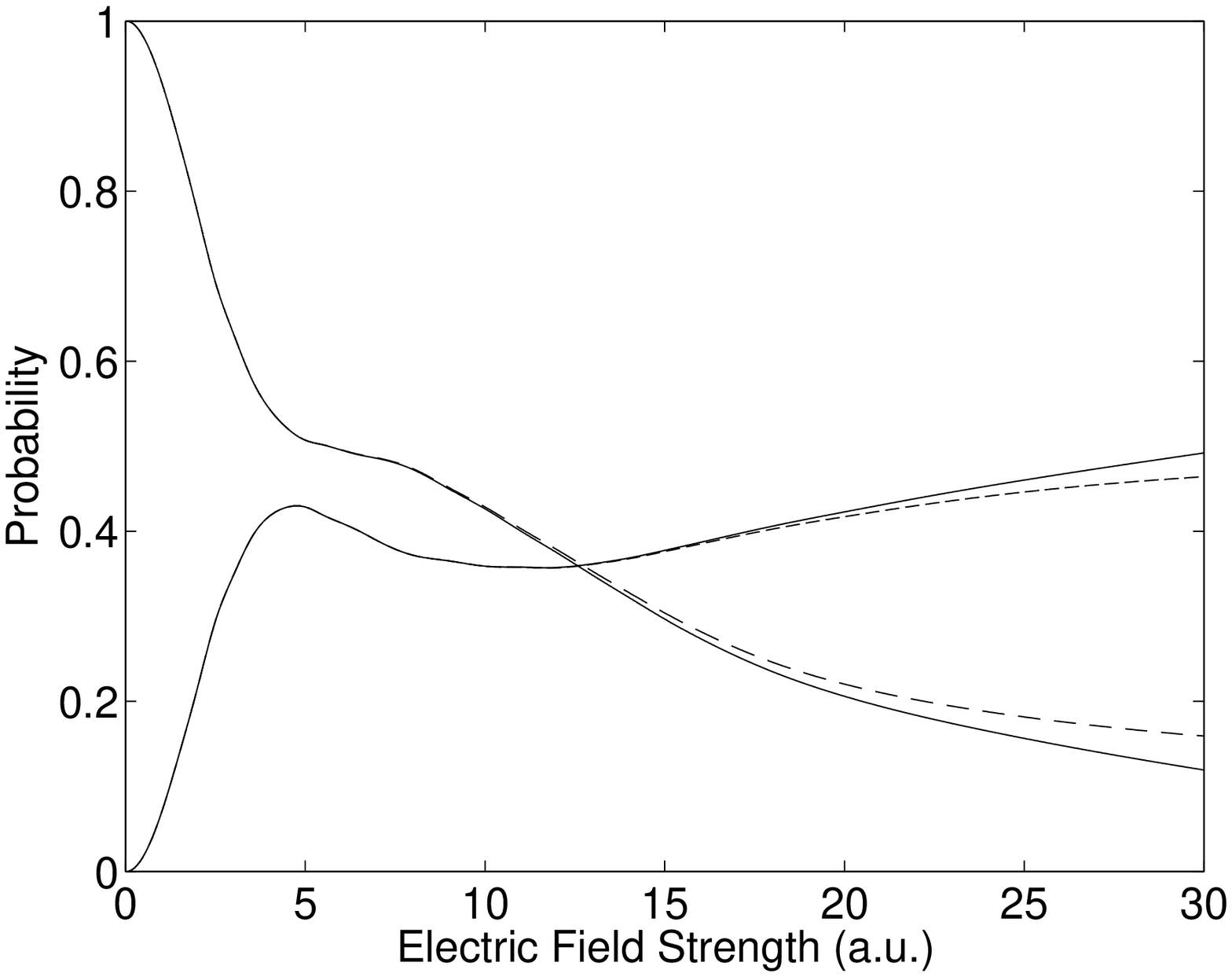}
\end{center}
\caption{\label{fig2} Upper panel: As Fig.~\ref{fig1}, but for the
fully three-dimensional case with the system initially prepared in
the H(1s) state. Lower panel: corresponding results for $\omega=2$
a.u.. }
\end{figure}
The wave function is expanded on the grid points $\left[(r_i,
\Omega_{jk}) =(r_i,\theta_j, \phi_k)\right] $ as
\begin{equation}\label{psi}
\Psi(r_i, \Omega_{jk},t) = \sum_{l,m}^{l_{max},m_{max}}
f_{l,m}(r_i,t)Y_{l,m}(\Omega_{jk}),
\end{equation}
and the initial field-free H(1s) state is obtained from the exact
analytical expression. Reflection at the edges $r=r_{max}=200$
a.u. is avoided by imposing an absorbing boundary. For
convergence, we include harmonics up to $l_{max} = 29$, check for
gauge invariance, use propagation time-step $\Delta t = 0.01$
a.u., and set $\Delta r= 0.2$ a.u.. Photoelectron probability
distributions are calculated by projecting the wave function onto
the field-free (discretized) continuum states. We note that the
presence of nondipole terms will lead to a population of different
$m$-values in (\ref{psi}).

In Fig.~\ref{fig2} total ionization- and ground state probabilities
are shown for the fully three-dimensional case in the nondipole and
dipole limits for two different frequencies.  We observe that the
dipole approximation remains valid up to field strengths of the
order of 10 a.u., and we find in general only a small effect of
the nondipole terms on stabilization.
\begin{figure}
\begin{center}
\epsfysize=5.5cm \epsfbox{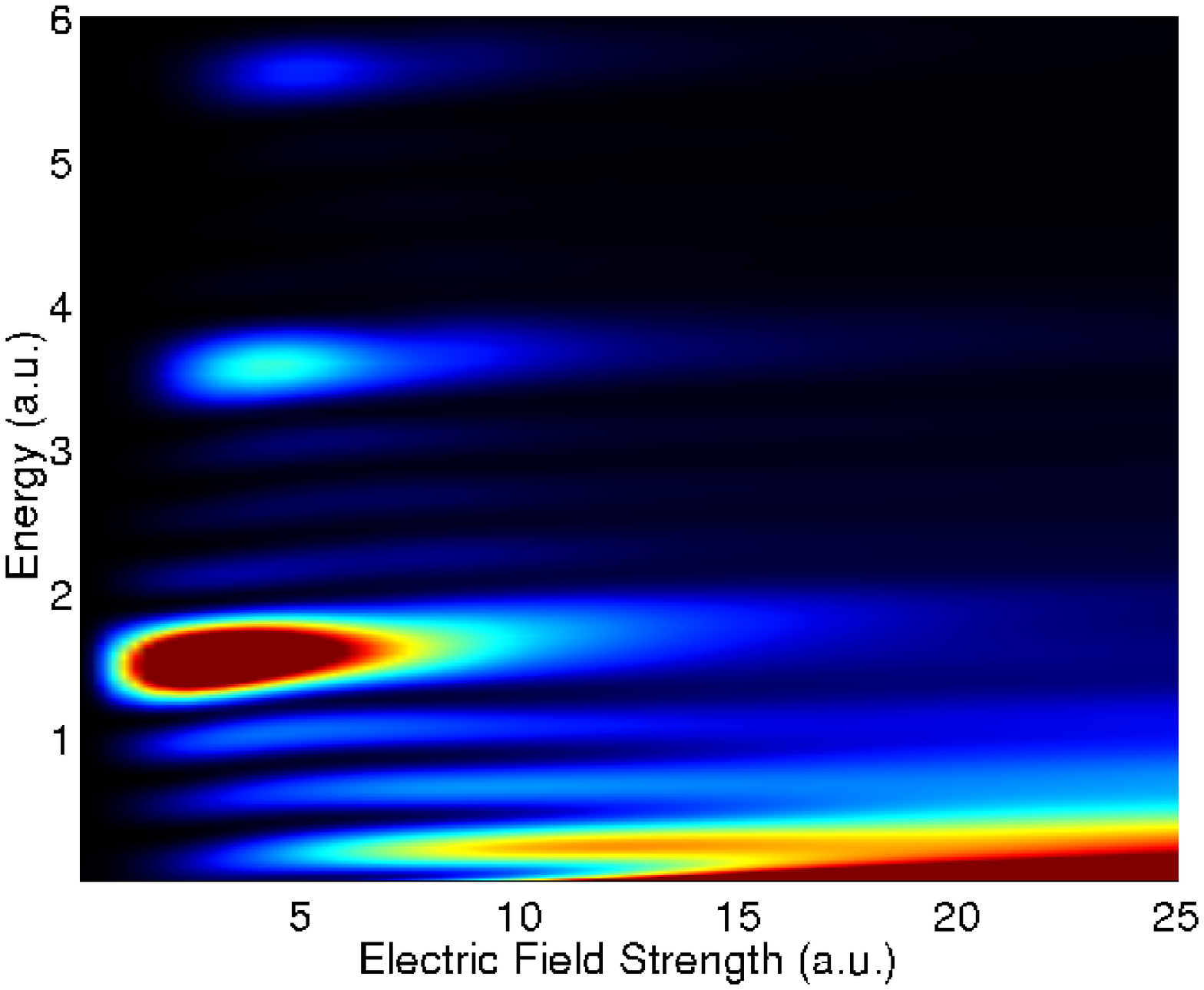} \epsfysize=5.5cm
\epsfbox{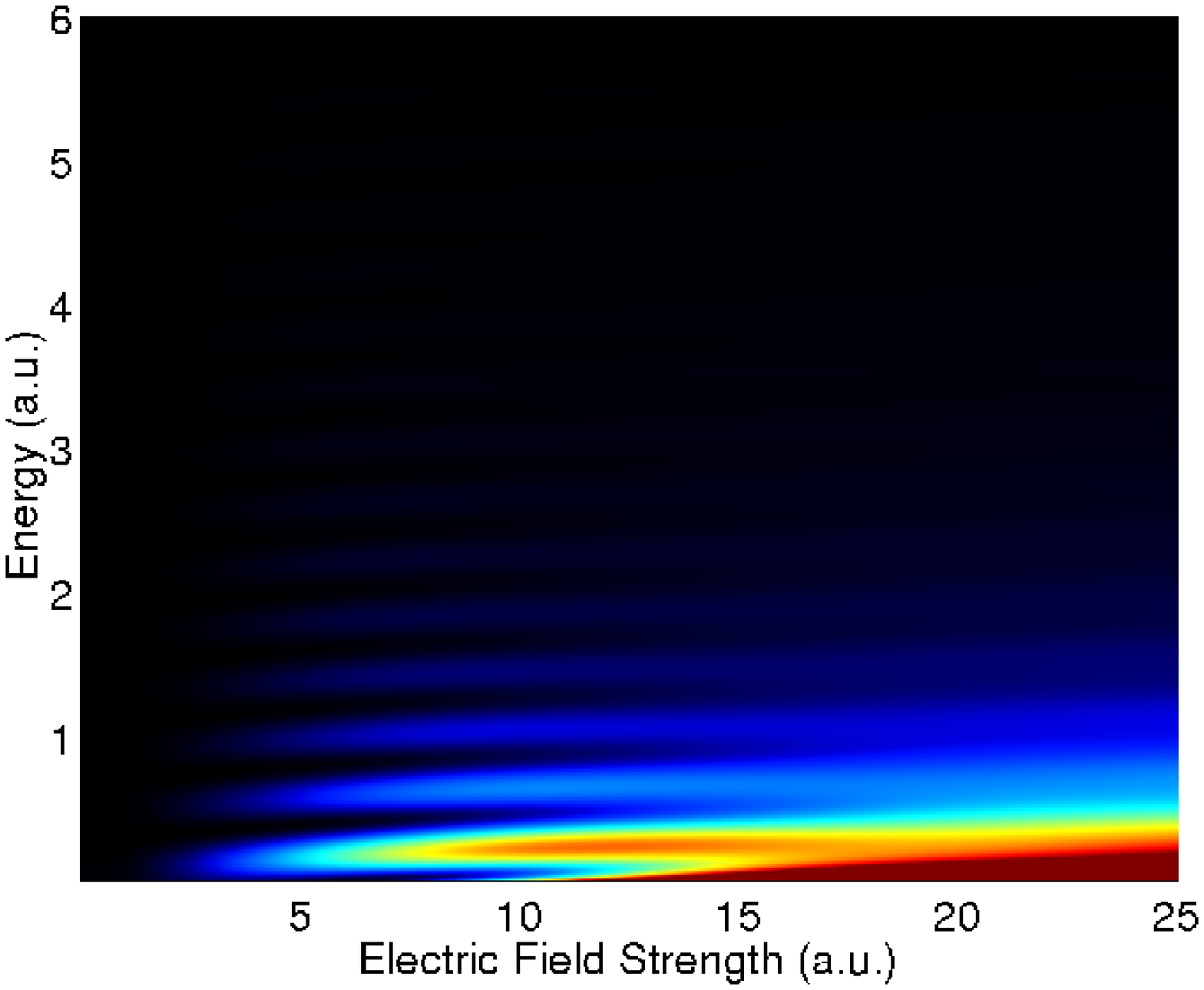} \epsfysize=5.5cm \epsfbox{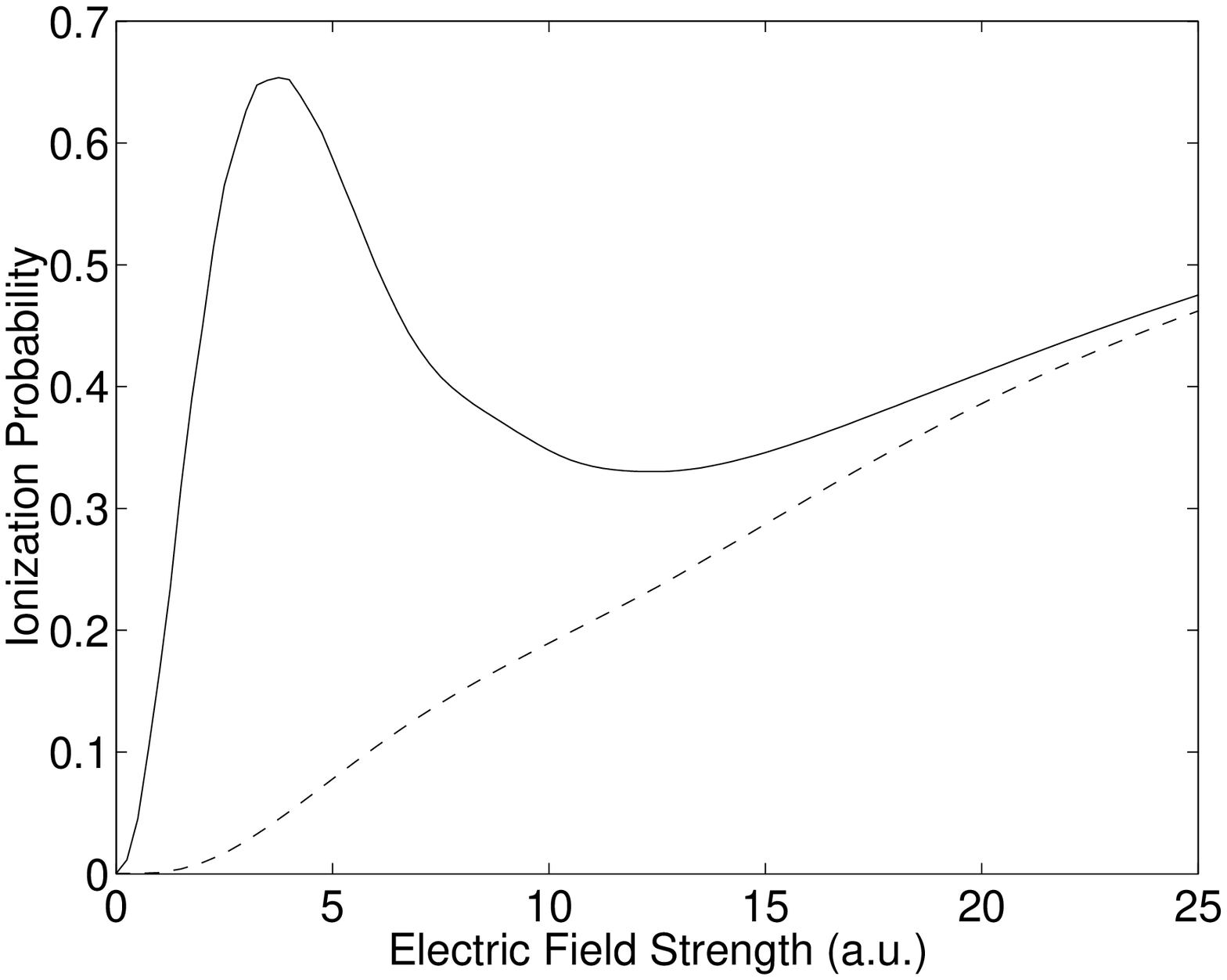}
\end{center}
\caption{\label{fig3} Upper panel: Ionization probability density
$dP/dE$ for H(1s)  vs.\ electric field strength and energy of the
ionized electron for a squared shaped 5-cycle pulse ($380$ as)
with $\omega=2$ a.u. ($23$ nm) by  fully three-dimensional
calculations. Middle panel: Results from the time-averaged
potential (\ref{lign4}). Lower panel: Total ionization probability
in the dipole approximation (full curve) and for the time-averaged
potential (\ref{lign4}) (dashed curve).}
\end{figure}

We now turn to the central question of how stabilization can most
efficiently be experimentally detected. Measurement of absolute
probabilities will require control of all parameters of the
experiment: atom density, pulse characteristics, repetition rates,
electron counts etc.. We therefore suggest measuring the
energy-differential photoelectron spectrum. Figure \ref{fig3}
shows the ionization probability density $dP/dE$ vs.\ electric
field strength and energy $E$ of the ionized electron with the
full interaction potential (upper panel) and with the
time-averaged Kramers-Henneberger potential (middle panel)
\cite{GavrilaPopov},
\begin{equation}
\label{lign4} V_0(\alpha_0;{\bm r})= \frac{1}{T}\int_0^T V({\bm
r}+\mbox{\boldmath $\alpha$})dt,
\end{equation}
where $\alpha_0\equiv E_0/\omega^2$ is the quiver amplitude. The
dipole and nondipole results are practically identical, and only
the dipole results are shown in Fig.~\ref{fig3}. For lower field
strengths a regular pattern of multiphoton resonances
corresponding to absorption of $1\omega$, $2\omega$ or $3\omega$
from the field is present. However, the multiphoton ionization
process weakens at higher intensities as the stabilization sets
in. Simultaneously, there is a steady growth in the portion of
low-energy photoelectrons in the spectrum which can be assigned to
$V_0$ of (\ref{lign4}). That $V_0$ is responsible for the growth
in the low-energy spectrum is readily seen by comparison of the
upper and middle panels. The processes leading to ionization
effectively divide into two competing classes: The multiphoton
ionization superimposed on a monotonically increasing 'background'
ionization process due to $V_0$ solely. This is explicitly
illustrated in the lower panel of Fig.~\ref{fig3}, where the total
ionization probability vs.\ electric field strength is
shown~\cite{comment}. Multiphoton ionization dominates at lower
field strengths, whereas the picture is the opposite at higher
values of $E_0$. The ionization due to the $V_0$ potential
reflects to what extent the laser pulse is turned- on and off
adiabatically, and in a 'sudden approximation' picture it
represents the lack of overlap between the field-free and the
field-dressed states. Common in both photoelectron spectra is the
presence of peaks in the probability density which cannot be
attributed to multiples of $\omega$. Instead, they are a result of
the non-adiabatic turn-on and turn-off of the field and can be
associated with the higher-order Fourier components of the pulse.

In summary, we presented a new formulation of the interaction
between atoms and light maintaining full spatial dependence of the
fields. We analyzed the terms in the interaction Hamiltonian and
argued, supported by numerical evidence, that certain terms can be
neglected. For the present field parameters, the main nondipole
effects come from $A(\eta)^2$. As an application, we considered
the phenomenon of dynamic stabilization in intense high-frequency
fields. We showed by full three-dimensional wavepacket simulations
that the nondipole terms do not destroy the stabilization effect,
and most importantly that the photoelectron spectra in the
stabilization regime shows very characteristic features: After
onset of stabilization all ionized electrons have very low kinetic
energy. Thus, by simply measuring the energy of the released
electrons stabilization can be detected.

\begin{acknowledgments}
It is a pleasure to thank Thomas K. Kjeldsen for useful
discussions and for critically reading the manuscript. The present
research was supported by the Norwegian Research Council through
the NANOMAT program and the Nordic Research Board NordForsk and by
the Danish Natural Science Research Council.
\end{acknowledgments}


\begin{thebibliography}{References}
\bibitem{comment}
Note that the difference between the full curve in the lower panel
of Fig.~\ref{fig3} and the ionization curve in Fig.~\ref{fig2} is
due to the different pulse forms used.
\end{thebibliography}
\end{document}